# Relaxation Time of Multipore Nanofluidic Memristors for Neuromorphic Applications

**Agustín Bou\*,[1] Patricio Ramirez,[2] Juan Bisquert[3]**

[1]Leibniz-Institute for Solid State and Materials Research Dresden, Helmholtzstraße 20, 01069 Dresden, Germany

[2]Dept. de Física Aplicada, Universitat Politècnica de València, E-46022 València, Spain

[3]Instituto de Tecnología Química (Universitat Politècnica de València-Consejo Superior de Investigaciones Científicas), Av. dels Tarongers, 46022, València, Spain.

*Email:* acatala@uji.es

**Abstract**

Memristors have been positioned at the forefront of the purposes for carrying out neuromorphic computation. Their tuneable conductivity properties enable the imitation of synaptic behaviour. Multipore nanofluidic memristors have shown their memristic properties and are candidate devices for liquid neuromorphic systems. Such properties are visible through an inductive hysteresis in the current-voltage sweeps, which is then confirmed by the inductive characteristics in impedance spectroscopy measurements. The dynamic behaviour of memristors is largely determined by a voltage-dependent relaxation time. Here, we obtain the kinetic relaxation time of a multipore nanofluidic memristor via its impedance spectra. We show that the behaviour of this characteristic of memristors is comparable to that of natural neural systems. Hence, we open a way to study the mimic of neuron characteristics by searching for memristors with the same kinetic times.

The rapid development of emerging memory technologies has fueled the search for novel materials and device architectures to overcome the limitations of traditional memory technologies.[1] Memristors, as a promising candidate in this field, are set to play a crucial role.[2,3] Their unique ability to retain a history of past voltages and currents, effectively functioning as a memory, offers a significant advantage in terms of power efficiency and speed.[4] These devices can be integrated into neuromorphic systems, which mimic the neural architectures of the human brain, paving the way for more advanced and efficient artificial intelligence applications.[5-7] With a growing demand of processing power due to the growth of such applications, the candidates for realizing memristic devices is increasing.[8,9]

Understanding the operation and performance of memristors is of crucial importance to implement such devices in real neuromorphic systems. However, given the variety of different material and architecture candidates for building memristic functional devices,[10-13] it is advantageous to have specific characteristics that permit inspecting the suitability of memristors for mimicking synaptic behaviour. Among these characteristics, we focus our research on their kinetic relaxation time. This element governs the speed of the transitions from one conductive state to another, and it is a crucial element of neuron models such as the Hodgkin-Huxley model.[14] The velocity of these transitions depends on the applied voltage and so it does the kinetic time. The analysis of this characteristic of memristors is of use when finding similarities with relaxation times of neuron models. When a memristor has a relaxation time with the adequate voltage dependence, it is suitable for its integration in neuromorphic systems. In addition, the relaxation determines the characteristics of switching times that usually present exponential dependence with the voltage in solid state memristors that contain ionic and electrochemical processes.[15-19]

The iontronic nanofluidic channels[20-24] have shown the characteristic inductive hysteresis of memristors, as well as the gradual tunable conductive potentiation desired for neuromorphic applications.[22,25-27] However, the relaxation time of these devices has not been yet explored. Here, we propose Impedance Spectroscopy (IS) as a potential tool for accessing this characteristic of memristors. Our approach suggests to look at the IS produced by neuronal systems and synapses exploiting the similarities in the IS experiments of memristors.[28] The key element for a neuromorphic applications of a memristor is the inductive low frequency arc,[24] and here we obtain the associated kinetic time.

Interestingly, there have been various reports using this technique, which have found similarities in the experimental IS spectra of memristors with those given by various neuron models. Specifically, the multipore nanofluidic memristors have already been reported to show IS spectra with sharp inductive elements.[29] Halide perovskite memristors have shown these type of inductive spectra, too, and they have been used to correlate the IS inductive behaviour with the hysteretic current-voltage curves characteristic of memristors.[30,31]

In a recent model, we have shown that both the inductive and the capacitive conduction effects can be described with a single relaxation equation producing both phenomena.[32] Therefore, we can conclude that both inductive and capacitive traces have the same physical origin. In fact, we simplify the equivalent circuit used for the IS analysis, and we can describe the full set of spectra with a single R-L branch.

This approach allows us to extract new information about the characteristics of memristors. Here, we apply our model to the analysis of multipore nanofluidic memristors with synaptic tunability.[25,33] We fit the experimental IS spectra of the memristor and we can obtain the characteristic time of the kinetic equation governing the conductivity, which depends on voltage. Both the elements and the kinetic time extracted from impedance have the expected voltage dependence of the model. With this approach, we lay the foundation for the effective correlation and integration of memristors with neuromorphic systems.

First, we present the previously developed model for the current-voltage $I_{tot} - u$ dynamic measurements, which consists of two equations[24,34,35]

$$I_{tot}(u) = [g_L + (g_H - g_L)x]u \tag{1}$$

$$\tau_k(u)\frac{dx}{dt} = x_{eq}(u) - x \tag{2}$$

Here, $g_L$ is the low conductance and $g_H$ the high conductance, see Fig. 1a, $x_{eq}$ is the equilibrium value of the memory variable $x$, and $\tau_k(u)$ is the relaxation time. We use the model functions[24]

$$x_{eq} = \frac{1}{1 + e^{-(u - V_{Bx})/V_m}} \tag{3}$$

$$\frac{1}{\tau_k(u)} = \frac{1}{\tau_M} + \frac{1}{\tau_0}\left(e^{\frac{u - V_A}{n_A V_m}} + e^{-\frac{u - V_D}{n_D V_m}}\right) \tag{4}$$

The variables $x_{eq}$ and $\tau_k$ are represented in Fig. 1c. Eq. (3) is a typical sigmoidal function with onset parameter $V_{Bx}$ and steepness $V_m$. We see the gradual change of $x_{eq}$ from the low conductive state, at negative voltages, to high conductive state $x_{eq} = 1$, at positive voltages. The relaxation time is described by the parameters $\tau_0, \tau_M, V_m, V_A, V_D, n_A, n_D$. It has two exponential wings and a central maximum value limited to $\tau_M$, see Fig. 1c.[24,34] This is a desired characteristic of memristors, since many biological channels, as in the Hodgkin-Huxley model,[14] as well as ionic/electrochemical solid state memristors,[15,36] present this characteristic.

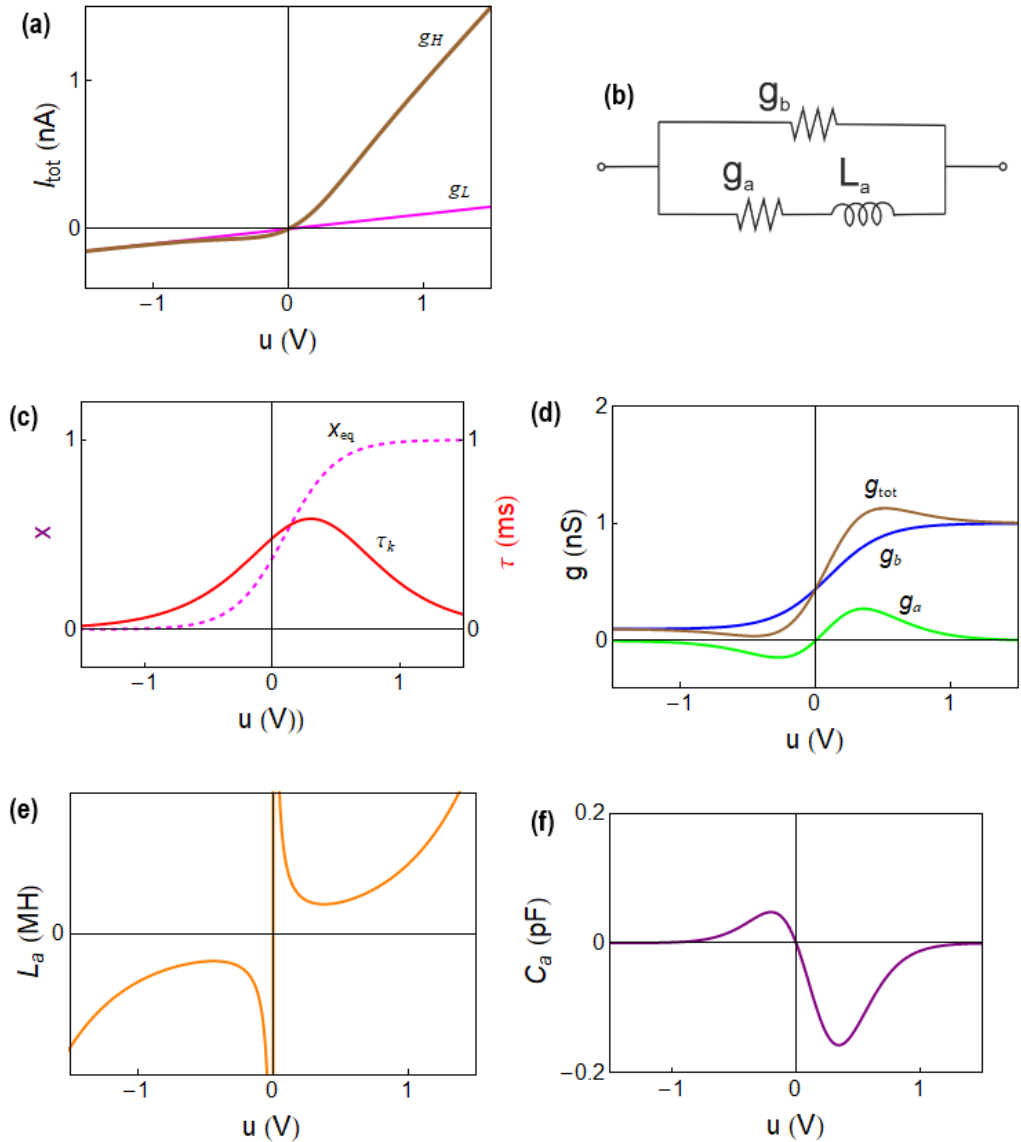

Fig. 1. Model representation. (a) Current-voltage curve. (b) Equivalent circuit. (c) Equilibrium value of the state variable and relaxation time. (d) Conductances. (e) Inductor. (f) Effective low frequency capacitance. The expressions of $x_{eq}$ and $\tau_k$ are given in SI. Parameters $g_L = 0.1, g_H = 1$, $V_m = 0.2, \tau_0 = 1, \tau_M = 2, n_A = n_D = 2$. $V_{Bx} = 0.1, V_{B\tau} = 0.3, V_A = 0.5, V_D = 0.1$. Time in ms, voltage in V, current in $n$A.

To obtain insight into the dynamical properties of the model, we calculate the small signal ac impedance response at the angular frequency $\omega$. As usual[37,38] the equations are expanded to the first order, where the perturbation of variable $y$ is indicated as $\hat{y}$, and the factor functions of each term are computed at equilibrium conditions. Furthermore, we transform the small signal equations to the frequency domain by the Laplace transform, $d/dt \rightarrow s$, where $s = i\omega$. We obtain the equations[34]

$$\hat{I}_{tot} = \left[ g_L + (g_H - g_L) x_{eq}(u) \right] \hat{u} + (g_H - g_L) u \, \hat{x} \tag{5}$$

$$\hat{x} = \frac{c_\mu}{1 + s\tau_k} \hat{u} \tag{6}$$

Here

$$c_\mu = \frac{dx_{eq}}{du} \tag{7}$$

plays the role of a chemical capacitance (here with dimension $V^{-1}$).[39] The solution of the impedance obtained from (5-7) is

$$Z(s) = \frac{\hat{u}}{\hat{\jmath}_{tot}} = \left[ g_b + \frac{g_a}{1 + s\tau_k} \right]^{-1} \tag{8}$$

The circuit elements are defined by the relationships

$$g_b = g_L + (g_H - g_L) x_{eq}(u) \tag{9}$$

$$g_a = c_\mu(u)(g_H - g_L)(u - E_0) \tag{10}$$

$$L_a = \frac{\tau_k}{g_a} \tag{11}$$

We also calculate the capacitance of the branch $(g_a, L_a)$ with impedance $Z_a = g_a^{-1} + sL_a$.

$$C_a = \frac{1}{s \, (g_a^{-1} + sL_a)} \tag{12}$$

Therefore

$$Re(C_a) = -\frac{g_a \tau_k}{1 + \tau_k^2 \omega^2} \tag{13}$$

In Figs. 1d and 1e we observe the conductivities and the inductor extracted from the model, as well as the total conductivity calculated from the sum of both conductivities. The conductivity $g_b$ shows a gradual change from a low conductive state to a high conductive state like that of the variable $x_{eq}$. We highlight the behavior of $g_a$ and $L_a$ which take both positive and negative values. Here, we observe the negative values at negative voltages, i.e. at the low conductive state, and positive values at positive voltages, i.e. at the high conductive state. We can see how the transition from negative to positive occurs at the transition region, as well as the kinetic time maximum.

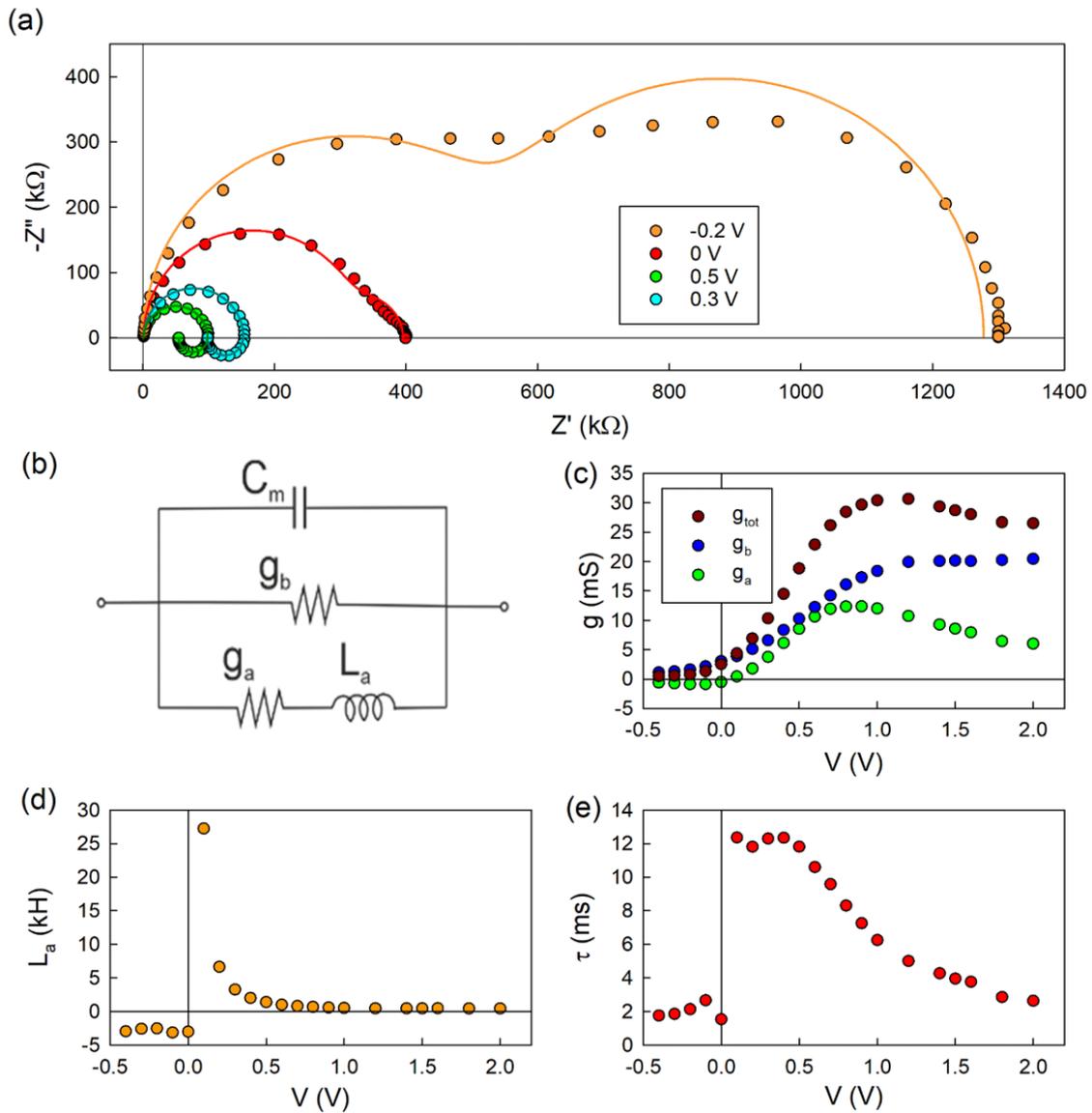

Fig 2. IS experimental data and analysis. (a) IS spectra at different voltages, points are the experimental data and straight lines are the fitting results. (b) Equivalent circuit used for the data fitting. Extracted parameters: (c) Conductance. (d) Inductance. (e) Relaxation time.

Fig. 2 shows both the IS experimental results of the multipore nanofluidic memristor and the IS analysis with the extracted parameters. In Fig. 2a, we show the experimental IS data, where we can observe the previously reported transition[31] from a two capacitive arcs spectrum at low conductive state or negative voltages, to an inductive low frequency arc spectrum at high conductive state or positive voltages. The equivalent circuit used for fitting the data is shown in Fig. 2b. Here, we have added a high frequency capacitor to that of Fig. 1b, in order to fit the high frequency arc present in all in membrane devices.

In Figs. 2b and 2c, we show the extracted elements from the fitted IS spectra. Interestingly, we find a satisfactory match between the expected behavior of the model conductances and the experimental ones. While $g_b$ takes only positive values transitioning from a low conductance to high conductance, $g_a$ changes from negative values to positive values at the transition region from low to high conductive state. Similarly to what we see in Fig. 1d, $g_{tot}$ takes its maximum at a certain voltage after the transition and then decays slightly reaching a plateau. In the case of the inductor we observe as well a change from positive to negative values, and although the shape of the experimental inductor does not follow exactly the trend of the model, it is similar in the way the transition takes place.

Finally, we can obtain the kinetic time from the elements of the experimental results in Fig. 2e. As expected from the model, we get a maximum of the characteristic time around the transition region, with two decay wings to lower (faster) time values at both sides of the maximum. The result satisfactorily matches both the model and the fact that the maximum hysteresis is reached around the transition region of the conductive state. This is the first time we can extract the kinetic time from an IS analysis, getting as a result a voltage dependence like those of natural neuron ion channels, leading to a prospective horizon in the integration of fluidic nanopore memristors in neuromorphic architectures.

To sum up, our study focuses on the potential of memristors for neuromorphic systems through the analysis of their IS spectra. We use IS as a tool to evaluate the suitability of asymmetric pore channels for mimicking synaptic behavior, identifying the key characteristic of an inductive low-frequency arc. Our experiments with multipore nanofluidic memristors show IS spectra with sharp inductive elements, matching neuron models. This indicates their potential for synaptic applications. The application of our model to experimental IS data reveals voltage-dependent characteristics consistent with theoretical predictions, including the kinetic time and conductance behavior. These findings validate the model and demonstrate the prospective integration of memristors into neuromorphic architectures, highlighting their potential for advanced memory and AI applications.

**Acknowledgements**

We thank the financial support from the Ministerio de Ciencia e Innovación of Spain (MICINN) project EUR2022-134045.